\documentclass[preprint,amsmath]{revtex4-1}

\usepackage[english]{babel}
\usepackage[utf8]{inputenc}
\usepackage{amsmath}
\usepackage{graphicx}
\usepackage[colorinlistoftodos]{todonotes}
\usepackage{xcolor}

\usepackage{blindtext}
\usepackage{epsfig}
\usepackage{amssymb}
\usepackage{color}
\usepackage[normalem]{ulem}
\usepackage{caption}
\usepackage{subcaption}
\usepackage{mathtools}
\usepackage{fixltx2e}
\usepackage{color}

\begin{document}
\title{The Effect of Disordered Substrate on Crystallization in 2D}

\author{Deborah Schwarcz}
\email{deborah.schwarcz@gmail.com}
\author{Stanislav Burov}
\email{stasbur@gmail.com}
\affiliation{Physics Department, Bar-Ilan University, Ramat Gan 5290002,
Israel}
\pacs{PACS}

\begin{abstract}
In this work the effect of amorphous substrate on crystallization is addressed. By performing Monte-Carlo simulations of solid on solid models we explore the effect of the disorder on crystal growth. The disorder is introduced via local geometry of the lattice, where local connectivity and transition rates are varied from site to site. A comparison to an ordered lattice is accomplished and for both, ordered and disordered substrates,  an optimal growth temperature is observed. Moreover, we find that under specific conditions the disordered substrate may have a beneficial effect on crystal growth, i.e., better crystallization as a direct consequence of the presence of disorder.

\end{abstract}

\maketitle

\section{introduction} %for title
Controlling self-organization of two dimensional crystals is very instrumental due to their potential applications in
low-dimensional semiconductors and optoelectronic devices \cite{Novoselov,Putkonen2005,Wu2019,Chen_19}.   For example, self-organization of a quantum dot \cite{Nurminen_1}, graphene sheet \cite{Tetlow_24} or  transition metal dichalcogenides \cite{Wu2019} affects their optical,  electrical, magnetic and mechanical
properties.  Two-dimensional crystals can be also assembled in three-dimensional hetero-structures that do not exist in nature and have tailored properties \cite{Novoselov}. %~\citep{epitaxial graphene 24, 25}%.

 This explains the considerable interest in studying metallic and semiconductor crystal growth. A promising technological development is to grow two-dimensional crystals on a substrate. The most studied two-dimensional crystal is graphene. Graphene has a wide range of interesting properties such as extremely large charge motilities, unprecedented mechanical strength, remarkable heat conduction at room temperature and a uniform absorption of light across the visible and near-infrared parts \cite{Novoselov}.  Growth rates of two-dimensional crystals, such as graphene are limited by surface diffusion or attachment, that is, by the rate at which deposited atoms jump towards the growing crystal or the rate at which these atoms attach to the crystal.  
Different parameters such as flux of atom deposition on the surface, temperature, and the nature of the substrate on which the crystal grows, influence the speed of crystal growth as well as the quality of the obtained crystal \cite{Wu2019,Levi_1997,Tetlow_24,Jourdain}.
Most theoretical studies that are aimed at providing a better understanding of a crystal growth have used Molecular Dynamics (MD) \cite{Meng_16,Tetlow_24,Shibuta} or Kinetic Monte Carlo (KMC) methods \cite{Tetlow_24,Nurminen_1,Zhang_17}. MD simulation (MDS) describes the diffusion of atoms according to the forces acting on them. The MDS approach involves the description of all the atomistic details \cite{Tetlow_24,Tang_25,Ratsch_26,Biehl} and crystal growth is associated with many local events. 
%Therefore, in order to describe the system, the integration time step must be small enough to capture the dynamics. This requires time steps in the femtosecond range. But the typical time between two jumps of an adatom can reach microseconds due to energy activation barriers. Therefore, if one attempts to run an MDS for a growth process in which numerous activation events are present, the system will spend most of the time sitting in the potential energy wells with extremely rare jumps between them \cite{Tetlow_24} . 
In this study we concentrate on KMC simulations, that in general, go beyond atomistic details. These methods are based on coarse-graining of  ``molecular" growth models. Coarse-grained methods take into consideration “important” processes and neglect other details  \cite{Tetlow_24, dBattaile, Chatterjee2007}. For example when describing crystallization on substrate one can look at jumps of absorbed atom (adatom) between neighbouring cites of immovable lattice that represent the substrate. %that move on a lattice. Adatom can jump to a neighboring cell, we assume that the substrate on which the crystal grow (the lattice in the simulation) not move. 
In reality, the substrate rearranges and reacts to  adatom movement but as long as the substrate movement has a negligible impact on the adatom energy, this movement can be neglected \cite{Biehl}.
This kind of KMCs describing crystal growth as adatom moving on a lattice, is called solid on solid (SOS) simulations \cite{Chatterjee2007,Pyziak,Biehl} . In SOS simulations, adatoms are deposed randomly on a site and then jump to a neighbor site according to a microscopic model (see Fig.~\ref{processes}).

The effects of substrate on nucleation and crystal growth have been previously investigated~\cite{Nurminen_1,Chen_19}  using SOS simulations. These models usually use a square lattice and add energetic barriers to specific sites. Specifically, in \cite{Nurminen_1} a patterned substrate was considered for the nucleation process. In order to introduce a substrate pattern, the lattice is divided into square-shaped domains and the adatom energy is a function of its location on the lattice. Nurminen et al.~\cite{Nurminen_1} found that patterned substrate affects the nucleation process on the substrate.  In \cite{Chen_19} point defects were introduced and it was found that this addition, at a certain  temperature of the substrate material, improves the nucleation and increases the average crystal size. 
While these studies show that patterning, or addition of local disorder to the substrate, have a good impact on crystal size (increased average island size) it is not clear what happens when the underlying lattice is not ordered, i.e., amorphous substrate.
%To conclude, it was  found that specific amount of defects \cite{Chen_19} or specially patterned substrate \cite{Nurminen_1} have a good impact on crystal size, meaning that the average island size is larger.  These models are useful for studies the defect impact on crystal growth but do not explain the case of amorphous substrate, in which the substrate is not ordered, in this case a square lattice is less appropriate. 
Moreover, experiments \cite{Celebi_8} show that substrate composition and surface crystallinity  also influence crystal growth. Recently, amorphous substrates such as liquid substrate, have been  experimentally used for crystal growth, the lack of a crystallographic substrate have been observed to have a good impact on crystallization, i.e. enlarge crystal size. \cite{Zeng_20,Boeck_27,Zhang_28}. When taking into account this experimental observation, it becomes interesting to explore the effect of substantial substrate disorder on crystal growth. Of special interest is the case when the substrate geometry is sufficiently altered and can't represent an ordered lattice anymore, i.e., the case of amorphous substrate.
%The geometry of the substrate is reflected by the lattice. 

The effects of temperature on crystal growth has been previously addressed in \cite{Meng_16,Zhang_17,Chen_19,Meixner_23,Pyziak}. In \cite{Meng_16} an optimal growth temperature for graphene growth on Ni is found by molecular dynamics simulations. An optimal temperature, at which the surface roughness is minimized,  was recently obtained~\cite{Zhang_17} using a 3D KMC simulation. %In this article, we present two SOS models and focus on the effects of temperature and substrate on crystal growth. 
Experiments show that at a relatively low temperature, temperature increase has a beneficial impact on crystal growth \cite{Celebi_8,Loginova_13,Hao_14,Chen_18} while, at higher temperatures, an opposite effect is found \cite{Rafik_9}.

In this work  crystallization is studied on two types of latices; square (ordered) and random (amorphous). We use a vectorizable random lattice (VRL) in order to simulate an amorphous substrate. VRL is a lattice with sites that compose a set of randomly chosen points with uniform distribution \cite{Moukarzel}. In \cite{Leonid} a VRL was used for the study of Lorentz gas. In \cite{Schwarcz_22} one of us used a VRL in order to simulate a semi-solid  substrate (agar substrate) that does not posses a crystal order.
Here we present two SOS models. Model A in which the interaction energy between adatoms depends only on the number of neighboring adatoms and model B where the interaction energy between adatoms depends on the local substrate geometry.
We compare the results obtained on both substrates, for both models, and the effect of temperature is also presented. Interestingly enough, it is found that not only that there is an optimal temperature, but also the amorphous nature of the substrate has a beneficial effect on crystallization.  
%The temperature, in contrast to the substrate geometry, appears in the equation model. High temperature lead to good adatom thermal diffusion, while low temperature decreases adatoms diffusion. 

This paper is organized as follows. In Sec. ~(\ref{model}) the models and computational methods are presented. The effects of temperature variation on crystallization and comparison of ordered and amorphous substrate are presented in Sec. ~(\ref{my result}). In Sec. ~(\ref{my result}) we also discuss the possible explanation for the beneficial impact of amorphous substrate. The summary is in Sec.~\ref{my discussion}.

\section {Models and methods }\label{model}

\begin{figure}
    \centering
    \begin{subfigure}[b]{0.28\textwidth}
        \includegraphics[width=\textwidth]{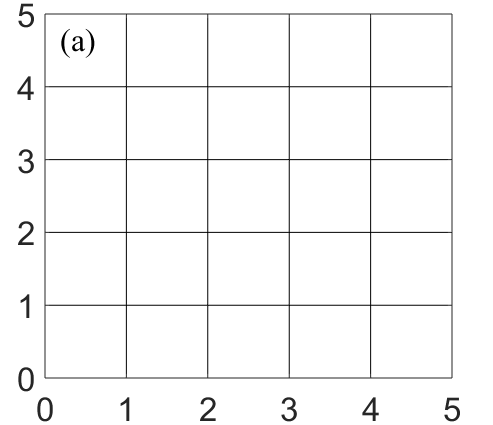}
        %\caption{square lattice definition}
    \end{subfigure}
    ~ %add desired spacing between images, e. g. ~, \quad, \qquad, \hfill etc. 
    \begin{subfigure}[b]{0.3\textwidth}
        \includegraphics[width=\textwidth]{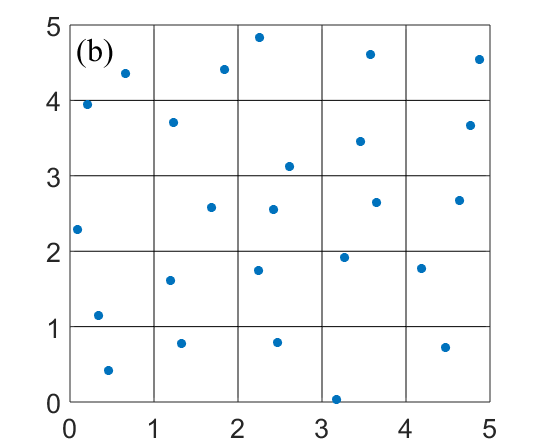}
     %   \caption{Random lattice site definition}
    \end{subfigure}
    ~ %add desired spacing between images, e. g. ~, \quad, \qquad, \hfill etc. 
    \begin{subfigure}[b]{0.3105\textwidth}
        \includegraphics[width=\textwidth]{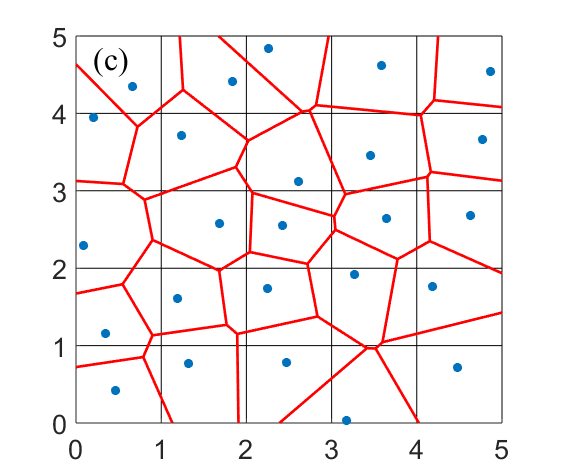}
        %\caption{Vornoi diagram of the lattice points}
    \end{subfigure}
    \caption{Vectorizable Random Lattice construction. {\bf (a)} Definition of a square lattice of $d\times d$ cells, the edge length of the reference square lattice is set to be $1$. {\bf(b)} Allocation of random lattice sites: at each cell of the ordered lattice a point is drawn with uniform distribution across the cell, while keeping a minimum distance $\delta$ between points in different ordered cells, in this figure $\delta$ is set to be $1/10$. {\bf(c)} Creation of Vornoi cells: Voronoi cell is defined as a set of points that are closer to a given lattice site than to any other lattice site. In this figure the Vornoi cells boundaries are drawn in red. Each Voronoi cell is a single cell of the VRL}
    \label{lattice}
\end{figure}

In order to describe crystal growth on an amorphous substrate by a  SOS model one first needs to create a lattice that represent an amorphous substrate. In order to achieve this goal we use a Vectorizable Random Lattice (VRL), VRL is a lattice that consists of random sites that are uniformly distributed in space    ~\cite{Moukarzel_21,Schwarcz_22}. During the construction of a VRL following steps are performed:  {\bf I}: A square lattice of $d \times d$ cells is defined, this lattice is called the reference lattice, as shown in Fig.~(\ref{lattice}) (a). {\bf II}: A random point is chosen in each cell of the reference lattice (with uniform distribution), while keeping a minimum distance $\delta$ between points as presented in Fig.~(\ref{lattice}) (b). Any two points that are closer than $\delta$, are re-allocated. $\delta$ controls the degree of randomness of the lattice.  These points constitute the VRL sites. {\bf III}: The Voronoi cell for each lattice site is produced Fig.~(\ref{lattice}) (c). Voronoi cell is defined as a set of all points that are closer to a given lattice site, than to any other lattice site~\cite{mixed-spin_IM_6}.
The simulations presented in this work were carried out on a reference lattice of $10^4$ cells ($d=100$), the edge length of the reference square lattice was set to be $1$ while $\delta$ is set at $1/10$. Special care was taken in order to enable periodic boundary conditions of the VRL. 
%Since the Voronoi diagram depends on the sites emplacement we follow the following steps in order to realize periodic boundary conditions.
%1. Add two lines/columns near each border.
%2. Copy to these lines/columns the relative emplacement of lattice sites of the opposite border.
%3. Draw the Voronoi diagram of the bigger lattice.
%4. Note the neighbor cell and other parameters (common border length for example) of the lattice cell. If a cell is bordering to lattice cells that was added in 1, we look at it's original cell. For example let's assume that site $k$ is on the first column of the lattice, therefor we will copy $k$'s relative emplacement (step 2), let's call this site $z$. If a site is bordering to $z$ it is actually bordering to $k$, because $z$ is a copy of $k$.   
%4. Return to the original lattice with the periodic boundary conditions.
In the following we describe the dynamics of adatoms on top of a VRL and an ordered square lattice.  
%%%%%%%%%%%%%%%%%%%%%%%%%%%%%%%%%%%%%%%%%%%%%%%%%%%%%%%%%
%%%%%%%%%%%%%%%%%%%%%%%%%%%%%%%%%%%%%%%%%%%%%%%%%%%%%%%%%

\begin{figure}
	\begin{center}
	\includegraphics[width=0.4\textwidth]{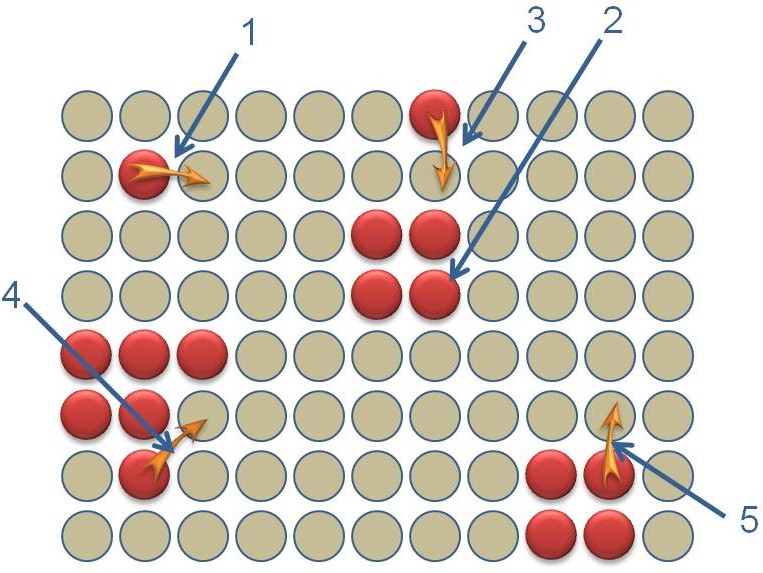}
		\caption{ Schematic representation of the 					processes included in the model. Red circles 				represent adatoms. 
	{\textbf{(1)}}  Adatom jumps to a nearest neighbor site.  {\textbf{(2)}}			Adatom nucleation, several adatoms assemble together and create a nucleus. .{\textbf{(3)}} Crysatl growth by attachment of a jumping adatom to a crystal.{\textbf{(4)}}Relaxation of an 			adatom that is a part of a crystal, i.e. in our model an adatom can modify it's emplacement in the crystal, especially if the adatom is located on the crystal border. .{\textbf{(5)}}				With relatively small probability adatom can detach from the crystal. 
	%These processes 	are described by our model.    
	}\label{processes}
	\end{center}
	
\end{figure}

Adatoms are allowed to move on the top of the disordered lattice, but first we  depose adatoms on the substrate i.e.  fill part of the lattice cells with adatoms. Therefore during the first 30 simulated time steps of the simulation 100 adatoms are randomly deposited on VRL sites. If the randomly chosen site is already occupied, the new  adatom is added to a neighboring site (in the case of random lattice the neighbor site with the bigger common edge is chosen). 
After this deposition period, adatoms are no further added. 
%After this time period of adatoms deposition,  adatoms are not added. 
During the simulation, each adatom can jump to one of the neighboring sites of the disordered lattice, i.e. sites that have a mutual edge with the current site that contains the adatom.   If an adatom jumps to a site adjacent to another adatom it nucleates to form a new crystal. If it jumps to a site adjacent to a crystal it aggregates to the existing crystal.  Adatoms that belong to a crystal border can detach from the crystal or change their emplacement in the crystal. From an energetic point of view, neighboring adatoms has interaction energy therefore if an adatom have a neighbor its jumping probability decreases. On the other hand the adatom probability to stay attached to its neighbors increases. Fig. ~(\ref{processes}) summarizes the dynamical process on the lattice, for simplicity the lattice is presented as symmetric. We allow at most one adatom per lattice site. 

%%%%%%%%%%%%%%%%%%%%%%%%%%%%%%%%%%%%%%%%%%%%%%%%%%%%%%%%%%
%%%%%%%%%%%%%%%%%%%%%%%%%%%%%%%%%%%%%%%%%%%%%%%%%%%%%%%%%%%%%%%%%%%%%%%

The probability of a randomly chosen adatom to jump is defined as a combination of two probabilities: the probability to leave the original site and the probability to occupy a new site. First, all the potential locations $x_j$s that the adatom may reach from its initial site $x_i$ are specified. These potential locations are the sites that have a common edge with $x_i$. For each of those sites the transition probability $p_{x_i,x_j}$ is calculated 
%%%%%%%%%%%%%%%%%%%%%%%%%%%%%%%%%%%
\begin{equation}
p_{x_i,x_j}=\frac{e^{\beta \Delta E_{x_i,x_j}}}{\sum_{j}e^{\beta \Delta E_{x_i,x_j}}},
\label{probability02}
\end{equation}
%%%%%%%%%%%%%%%%%%%%%%%%%%%%%%%%%%%
%as define in Eq. ~(\ref{probability02}).
where  $\beta=\frac{1}{k_b T}$, $E_{x_i}$   is the adatom energy in position $x_i$, $\Delta E_{x_i,x_j}=E_{x_j}-E_{x_i}$, T is the temperature and $k_B$ is the Boltzmann constant.
A destination site is randomly chosen in accordance to $p_{x_i,x_j}$.
The probability, $p_{x_i}$, that an adatom will attempt to leave its position $x_i$ is defined as
\begin{equation}
p_{x_i} =e^{-\beta E_{x_i}}.
\label{Arrhenius}
\end{equation}
Where $E_{x_i}>0$ is the adatom energy. $E_{x_i}$ is defined as a sum of two terms $E_{Int_i}$ and $E_{S_i}$ that represent the interaction with the substrate ($E_{S_i}$), and the interaction energy contribution ($E_{Int_i}$) from all of the occupied nearest neighbors.
%\begin{equation}
%E_{x_i}=E_{S_i} + E_{Int_i}.
%\end{equation}

The substrate binding energy, $E_{S_i}$,  is defined in direct proportion to the cell boundary size
%%%%%%%%%%%%%%%%%%%%%%%%%%%%%%%
\begin{equation}
E_{S_i}=\left(i_{th}\; cell\; perimeter\right)/a
\label{substrate energy}
\end{equation}
%%%%%%%%%%%%%%%%%%%%%%%%%%%%%%%%%%%%%%%%%%
where $a$ is a numerical coefficient. 
%that correlates between the order of magnitude of the adatom-substrate binding energy and the binding energy between adatoms. In the simulations we use the value $a=10$. In other words, without $a$, the interaction energy between adatoms is neglectable in comparison to $E_{S_i}$. 
$E_{S_i}$ is the local energetic well that dictates the diffusion properties of an isolated adatom. It depends on the lattice site geometry/size and is quenched, i.e. remains constant during the simulation. When the lattice is geometrically disordered, $E_{S_i}$ is responsible for introduction of this heterogeneity into the dynamics of the adatoms. In the simple case of an ordered lattice, the substrate binding energy $E_{S_i}$ is uniform for all the lattice sites and local diffusion is homogeneous without preferred locations.  Cases with uniform quenched energetic barriers were studied in \cite{michael, Dongwei}.
In our model we assume that the substrate irregularity also affects adatom emplacement and that the adatom emplacement, in its turn,  affects the interaction energy between adatoms. The definition of the interaction energy, $E_{Int_i}$, between the adatom $x_i$ and neighbouring adatoms $\{x_j\}$ is varied between the two models. In model A, for each cell every neighboring adatom has the same contribution to the interaction energy. In model B the interaction energy between adatoms depends on the local substrate geometry, specifically it depends on their common edge length $f_{x_i,x_j}$.
%To be precise in model B the interaction energy between adatoms depends on their common length edge $f_{x_i,x_j}$. 
The interaction energy between two neighboring adatoms in Model A is defined as 
\begin{equation}
E _{Int_i}=\frac{\sum_{j}b_{x_j}}{\sum_{j}1}
\label{Model_A_energy}
\end{equation}
where $b_{x_j}$ is $1$ if $x_j$ is occupied and $0$ otherwise, the sumation is over all the neighbouring sites.
% \begin{equation*}
%   b_{x_j}=
%    \begin{cases}
%      1,&  \mbox{ j is full} \\
%      0,&  \mbox{ j is empty}
%    \end{cases}
%\end{equation*}
In model (B) the binding energy between neighboring adatoms is 
%%%%%%%%%%%%%%%%%%%%%%%%%%%%%%%
\begin{equation}
E_{Int_i}=\frac{\sum_{j}b_{x_j} f_{x_i,x_j}}{\sum_{j}f_{x_i,x_j}}
\label{Model_B_energy}
\end{equation}
%%%%%%%%%%%%%%%%%%%%%%%%%%%%%%%%
where $f_{i,j}$ (as previously mentioned) is the length of the boundary between $x_i$ and $x_j$. The summation is again over all neighbouring sites.  In the specific case when an ordered lattice is used, Model A and B should provide exactly the same results. For a VRL there is expected to be a difference between the behavior of Model A and Model B, due to the fact that the boundaries of a Voronoi cell are random.  VRL simulates an amorphous substrate, therefore for a VRLs the substrate-adatom interaction energy $E_{S_i}$ varies for different cells.  In model A, for each cell there is a constant interaction energy between adatoms and their neighbors. For VRL the amount of neighbours for different cells is not uniform, this in turn cause to variations of $E_{Int_i}$ simply due to the fact that the summation in Eq. ~(\ref{Model_A_energy}) varies. In model B, on top of the diversification of Model A, the magnitude of interaction between two adatoms also varies according to $f_{x_i,x_j}$ ( Eq. (\ref{Model_B_energy})).
The reasoning behind introduction of this specific dependence of interaction energy between adatoms stems from the geometrical properties of the VRL. We want to take into account only the interaction with nearest neighbours. Number of nearest neighbours varies from site to site due to the disordered structure of the VRL. While for Model A the only thing that matters is the occupation of a nearest neighbour by adatom, in Model B the length of the mutual boundary with this specific site (i.e. $f_{i,j}$) is taken into account. The closer the centers of two VRL sites $i$ and $j$ the larger (on average) is their mutual boundary and vice versa. 

%In the case of square lattice, both models led to the same behavior, but for random lattice the difference between the models is significant because in the case of random lattice,  lattice cells are not identical to each other.
Both models include three scaling parameters, $a$, $\delta$ and the edge length of the square lattice. Theses parameters can be adjusted to correspond to
a given disordered situation and comparability between $E_{S_i}$ and $E_{Int_i}$ . The order of magnitude of $f_{i,j}$ is dictated by the edge length of the reference square lattice and $\delta$. When the length edge of the reference square lattice is one, increasing $\delta$ from $\epsilon (\to 0)$ to 0.3 leads to decrease of the variance of $f_{i,j}$. 
The distribution of $f_{i,j}$ is rather complicated and for high enough values of $\delta$ ($\delta\gtrapprox 0.5$) it stops to behave as a Gaussian centerd around its mean.
%Further increase of $\delta$ causes to a split of $f_{i,j}$ mean value, 
Therefore we chose to use rather small $\delta=0.1$. The  numerical  coefficient $a$ that appear in Eq. \ref{substrate energy} dictates the ratio of the  interaction energy of adatoms ($E_{Int_i}$) and the interaction energy between the substrate and the adatom ($E_{S_i}$).

To summarize, at each time step a particle is chosen in a random fashion, then its probability to jump, $p_{x_i}$ , is calculated in accordance to Eq. (\ref{Arrhenius}). If the  jump move is accepted, the destination site is chosen among all its empty neighbouring cells. An empty neighbour cell with a lower energy state, i.e. a neighbour surrounded by more adatoms, has better chance to receive the jumping adatom, in accordance to $p_{x_i,x_j}$. The probability $p_{x_i}$ (Eq. (\ref{Arrhenius})) of an adatom to exit from its original site,  match to trap models where the depth of the trap depends on the energy of the original site \cite{BOUCHAUD, stas2011,Akimoto,stas2017,Magdziarz2018}. The probability to reach a new site, as defined in Eq.  (\ref{probability02}),  resembles barrier models \cite{David_chandler_chapter, BOUCHAUD}, where the transition probabilities between neighbouring sites are asymmetric.

\section {Results} \label{my result}

In the simulation, crystallization on a substrate is reproduced by a two-dimension SOS model described in the previous section. The largest and the most defect-free crystals are the most desirable. In order to quantify the quality of a crystal, we need to take into consideration the crystal size and its uniformity. For this purpose we define an order parameter, the normalized weighted density (NWD): a sum over all cell edges that separate two occupied cells, normalized by the sum of all edges for all occupied cells,
\begin{equation}
NWD=\frac{\sum_{i} \sum_{j} b_{x_j} f_{x_i,x_j}}{\sum_{i} \sum_{j}f_{x_i,x_j} }
\label{NWD}
\end{equation}
where \begin{equation*}
   b_{x_j}=
    \begin{cases}
      1,&  \mbox{ j is full} \\
      0,&  \mbox{ j is empty}
    \end{cases}
\end{equation*}
The first summation (over $i$) is over all lattice cells in which there is an adatom, second summation (over $j$) is over the neighbors of cell $x_i$.
A brief glance at NWD reveals that there is a resemblance between NWD and the interaction energy of all the adatoms. The reason for this resemblance lays in the fact that the adatoms interaction energy has a strong effect on crystal growth.  A global perspective of the system shows that NWD increases with the compactness of a crystal:  
the higher is the number of occupied sites of a given adatom, the higher is NWD.
%\iffalse as presented in Fig. ~(\ref{NWD})b. \fi 
In the case of irregular crystal morphology, such as protrusion, branches and pores, many adatoms have empty neighbors and NWD is expected to be relatively small. 
%\iffalse as show in Fig. ~(\ref{NWD})a and Fig. ~(\ref{NWD})c. \fi

Two reasons can cause NWD to be small; if the adatoms jumping probability $p_{x_i}$ is highly restricted, the adatoms will be stuck in an isolated site, if $p_{x_i}$ is too high  adatoms will not stabilize even on sites with many neighboring adatoms.

\subsection {Temperature effects}
\begin{figure} 
    \centering
    \begin{subfigure}[b]{0.48\textwidth}
        \includegraphics[width=\textwidth]{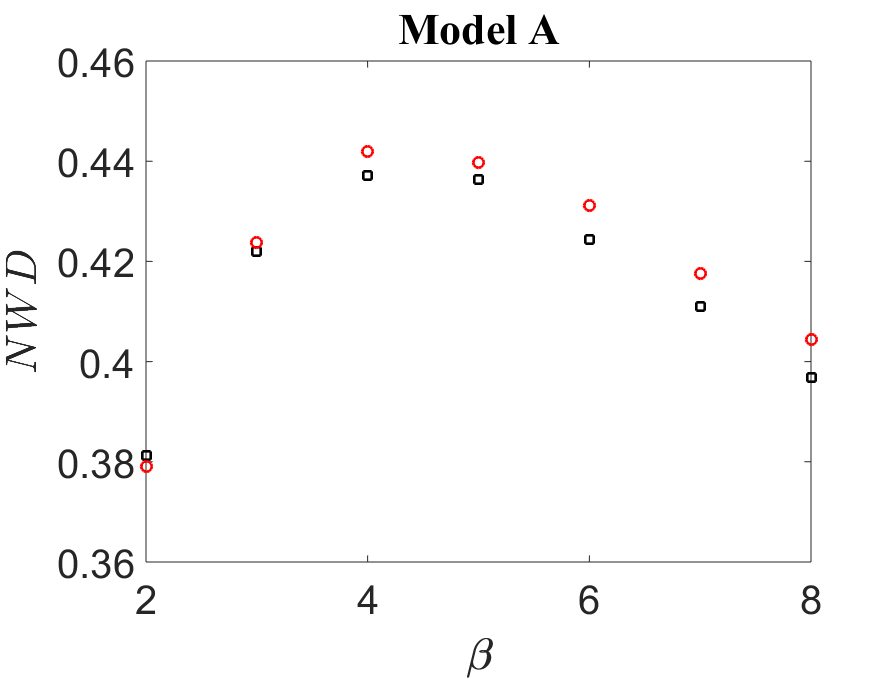}
        %\caption{} 
        \label{NWD_model_A_time_500}
           \end{subfigure}
 ~ %add desired spacing between images, e. g. ~, \quad, \qquad, \hfill etc. 
    \begin{subfigure}[b]{0.46\textwidth}
        \includegraphics[width=\textwidth]{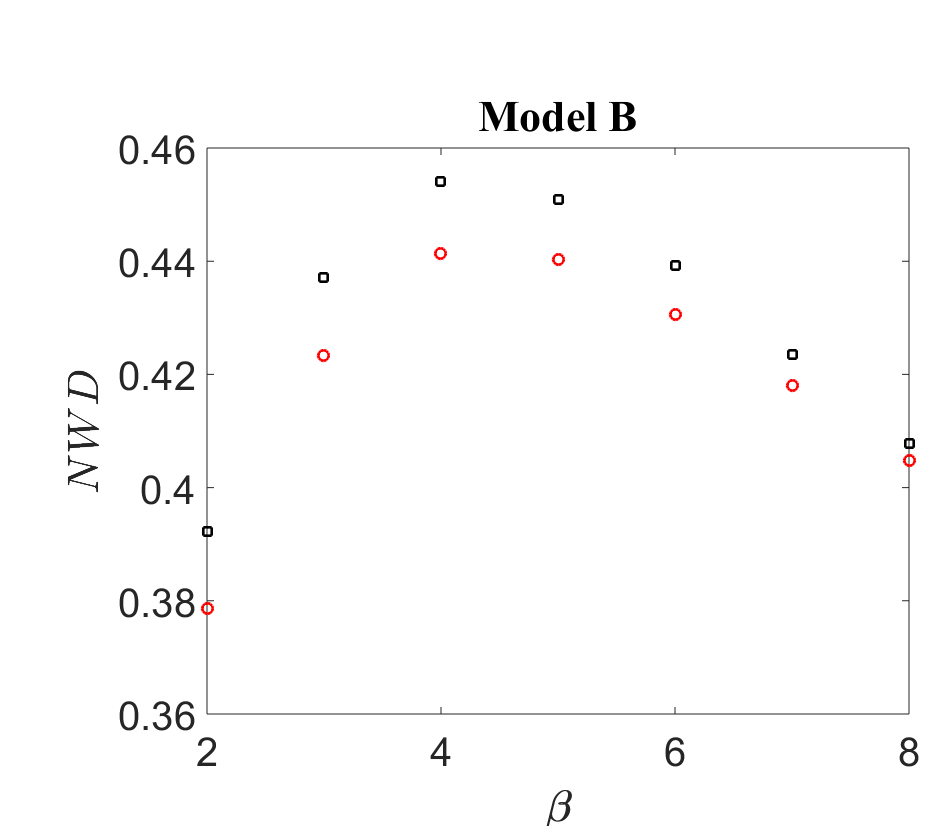}
        %\caption{} 
         \label{NWD_model_B_time_500}
    \end{subfigure}
 ~ %add desired spacing between images, e. g. ~, \quad, \qquad, \hfill etc. 
    \caption{NWD dependence on temperature for Model A ({\bf (a)}) and Model B ({\bf (b)}) and comparison to NWD for the ordered case of square lattice. Squares (\textcolor{black}{$\square$}) represent the behavior for disordered substrate, i.e. VRL, while circles (\textcolor{red}{$\bigcirc$}) represent the ordered square lattice. The computation of NWD was performed after $6\times 10^4$ simulation time steps. Averaging over $100$ realizations was performed, the error bar are smaller than the size of the presented symbols. 
    %, after 50,000 simulation time steps, red points represent the result over a square lattice, black points represents the result over a random lattice. the presented results are the averaged over 100 simulations, the error bar are smaller then the point  therefore we don't show them. (a) Model A. (b) Model B.
    }
    \label{littel_time_NWD}
\end{figure}

\begin{figure}
  \begin{subfigure}[b]{0.48\linewidth}
    \centering
\includegraphics[width=\textwidth]{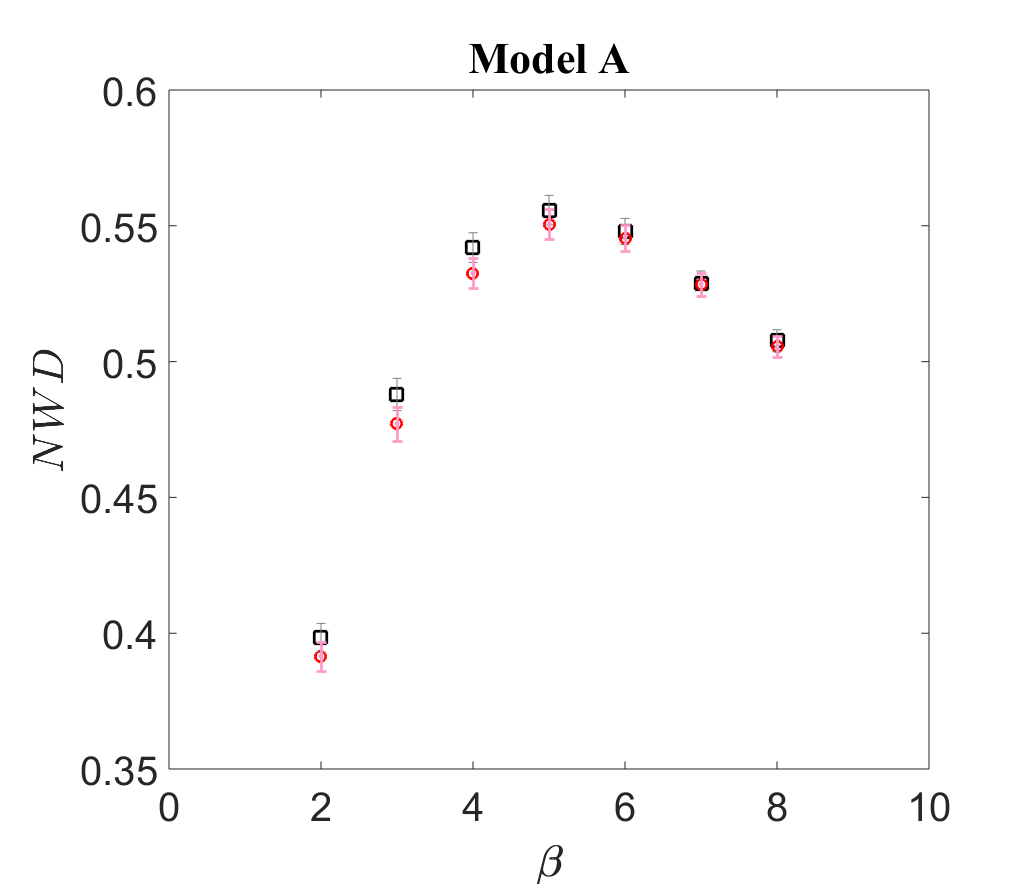}
        %\caption{} 
        \label{model_A_big_time}
  \end{subfigure}%%
  \begin{subfigure}[b]{0.47\linewidth}
    \centering
    \includegraphics[width=\textwidth]{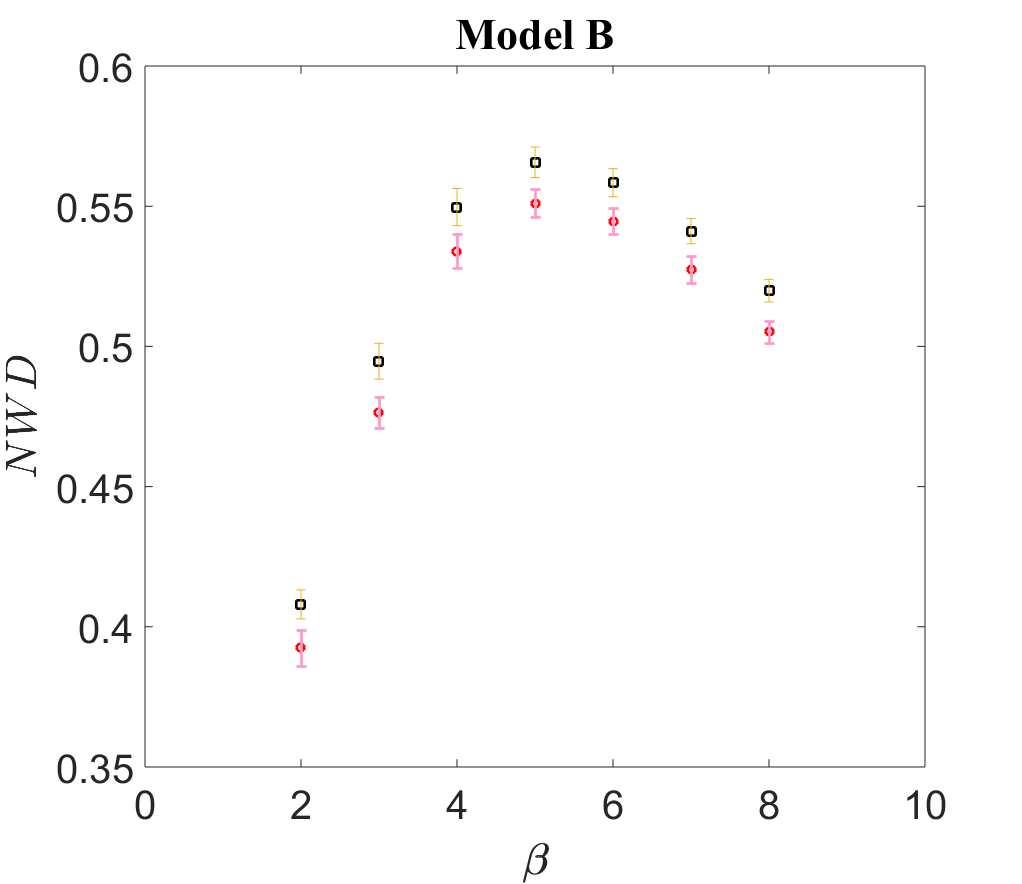}
       % \caption{} 
        \label{model_B_big_time}
  \end{subfigure} 
  
  \caption{ NWD dependence on temperature for Model A ({\bf (a)}) and Model B ({\bf (b)}) and comparison to NWD for the ordered case of square lattice. Squares (\textcolor{black}{$\square$}) represent the behavior for disordered substrate, i.e. VRL, while circles (\textcolor{red}{$\bigcirc$}) represent the ordered square lattice. The computation of NWD was performed after $6\times 10^5$ simulation time steps. Averaging over $100$ realizations was performed and error bars are presented.}
  \label{model_A_twist} 
\end{figure}

In order to study the effect of temperature, we vary  $\beta$ which appears in Eqs.~ (\ref{probability02})~and~(\ref{Arrhenius}). We observe that the effect of temperature on NWD has an inverse U shape, see (Fig.~(\ref{littel_time_NWD} (\textbf{a},\textbf{b}) and \ref{model_A_twist} (\textbf{a},\textbf{b}) ). At sufficently low $\beta$ (high temperatures) ,an increases of  $\beta$ improves crystallization until an optimal $\beta$ is reached, further increase of $\beta$ (low temperatures) damages the crystal. 

This observation of optimal temperature for crystal growth can be explained by following reasoning.
At sufficiently low temperatures the adatoms mobility is restricted due to the fact that they need quite large activation energy in order to leave their current site, i.e. small $p_{x_i}$. Increasing of the temperature leads to increasing of $p_{x_i}$ that in its turn contributes to better crystallization, i.e., higher NWD. 
%The reason that under a given temperature threshold crystallization is improved with increaseing the temperature is due to the fact that at low-temperature adatoms are stuck since an adatom doesn’t have the required activation energy to excite its initial position i.e.  $p_{x_i}$ in Eq. ~(\ref{Arrhenius}) is  small. Increasing temperature will increase $p_{x_i}$ and cause the adatom to be more mobile and thus it has a larger probability to reach a site with more neighbors. 
As the temperature is further increased, another aspect of the dynamics must be taken into account.
Above a specific temperature threshold 
%, $p_{x_i}$ is high enough and the chosen adatom will jump, but in this case, 
the difference between a neighbor site with many neighbors and a less favorable neighbor is small. From the Eq. ~(\ref{probability02}), we can see that $\beta$ decrease (T increase) causes a decrease in $\Delta E_{x_i,x_j}$ impact, therefore, the probability that a diffusing adatom will reach the site with highest interaction energy declines when the temperature is increased. This behavior is reproduced in our model (Figs.~\ref{littel_time_NWD}(\textbf{a},\textbf{b}) ,  \ref{model_A_twist}(\textbf{a},\textbf{b}) ) and was already observed  in \cite{Luis}.  
An optimal temperature for crystal growth was also observed experimentally by \cite{Rafik_9} and reproduced theoretically \cite{Meng_16,Zhang_17,Chen_19}.

\begin{figure} 
  \begin{subfigure}[b]{0.47\textwidth}
    \centering
            \includegraphics[width=\textwidth]{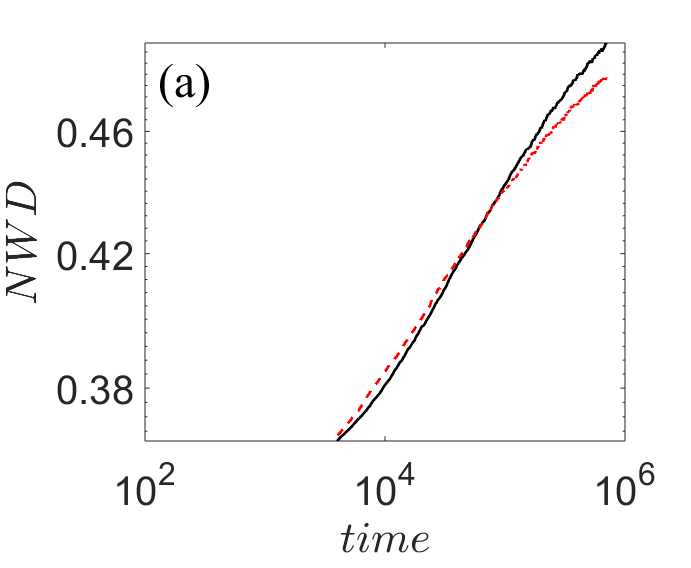}
        %\caption{} 
        \label{example_beta_3}
           \end{subfigure} 
  \begin{subfigure}[b]{0.49\linewidth}
    \centering
    \includegraphics[width=\textwidth]{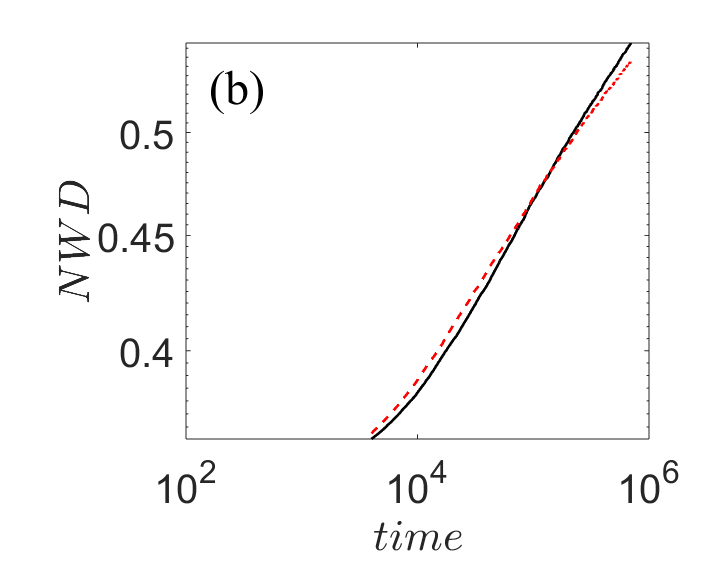}
        %\caption{} 
        \label{example_beta_4}
  \end{subfigure} 
 
  \begin{subfigure}[b]{0.48\textwidth}
    \centering
            \includegraphics[width=\textwidth]{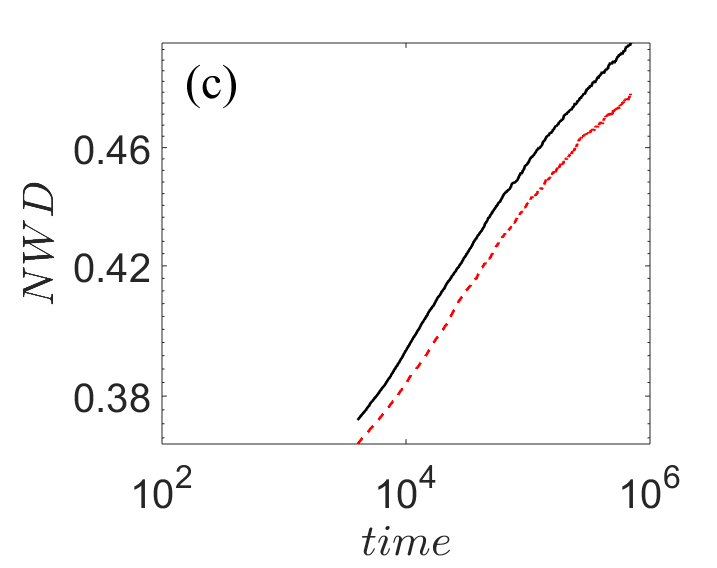}
        %\caption{} 
        \label{example_beta_3_model_B}
           \end{subfigure} 
  \begin{subfigure}[b]{0.48\linewidth}
    \centering
    \includegraphics[width=\textwidth]{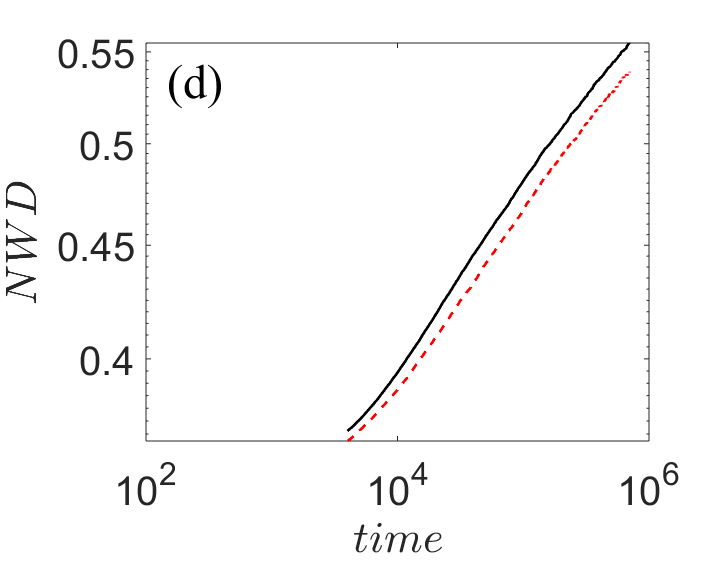}
        %\caption{} 
        \label{example_beta_4_model_B}
  \end{subfigure} 
  
  \caption{
  NWD growth with time for the disordered and ordered cases. Panel {\bf (a)} displays Model A (thick line) and square lattice (dashed line) behavior for $\beta=3$ while panel {\bf (b)} is for $\beta=4$.
  Panel {\bf (c)} displays Model B (thick line) and square lattice (dashed line) behavior for $\beta=3$ while panel {\bf (b)} is for $\beta=4$. Time is presented in simulation steps. The averaging was performed over $100$ realizations.
 % Model A NWD in dependence of time for (a)$\beta$=3 (b)$\beta=4$.
 %Model B NWD in dependence of time for (c)$\beta$=3 (d)$\beta=4$.
 }
 \label{NWD_vs_time}
  \end{figure}

\subsection{Effect of the Disorder}

The NWD is also used in order to study the effect of the disorder, i.e., amorphous lattice described by the VRL. For both definitions of the interaction energy, Model A and Model B, the comparison between ordered (square) substrate lattice and VRL with $\delta=0.1$ was performed ( Figs.~\ref{littel_time_NWD}(\textbf{a},\textbf{b}) and \ref{model_A_twist}(\textbf{a},\textbf{b})).  In Figs.~(\ref{littel_time_NWD} (\textbf{a},\textbf{b})) and (\ref{model_A_twist}(\textbf{a},\textbf{b})) the behavior of NWD for ordered and disordered cases is presented. Surprisingly enough, the effect of the disorder on crystal-growth is non-negative and can also be positive.

Specifically, for Model B, the NWD parameter is  higher for the amorphous case, for any given temperature. The growth of the crystal with time, as described by NWD, is slow in both cases (ordered and disordered) and  after sufficient temporal period appear to grow logorithmically (Fig.~\ref{NWD_vs_time}(\textbf{a}-\textbf{d})). 
The growth on the disordered substrate appears to be a little bit faster, for Model B, when compared to the ordered substrate. 
When Model A is considered the situation is somewhat more complicated. In Fig.~(\ref{littel_time_NWD} ({\textbf{a}})), there is a small preference for the ordered substrates for most temperatures. But for longer times, Fig.~(\ref{model_A_twist}(\textbf{a})),this seems to change and for sufficiently low  temperatures the disordered substrate starts to become more beneficial, as compared to the squared lattice. Indeed, when comparing the growth with time of NWD Fig.~(\ref{NWD_vs_time} (\textbf{a},\textbf{b})), we see crossover between the ordered and disordered substrates. In both cases the growth is logarithmic but the growth for the disordered case is faster.  

%Comparison between the NWDs of a square lattice and the NWDs of a random lattice at \textbf {relatively short simulation time}, reveal that when the interaction energy between adatoms depends only on the neighbors number, (model A) the square lattice lead to better NWD, as show in Fig. \ref{NWD_model_A_time_500}. While when the length edge between adatoms cells influence the interaction energy magnitude (model B) the crystals are more compact in the case of random lattice, as observe in  Fig. \ref{NWD_model_B_time_500}.

%At \textbf{long period} the random lattice led to better results also for model A as observed in Fig  \ref{model_A_twist}. 

%from here

 We must note that our results describe the stage when the crystal is still growing and local deformations are still in place. In Fig.~(\ref{NWD_vs_time} (\textbf{a},\textbf{b})), even after $\approx 10^6$ simulation time steps, the steady state, i.e. constant NWD, is not  reached.

\begin{figure} 
\centering
			\includegraphics[width=1.0\textwidth]{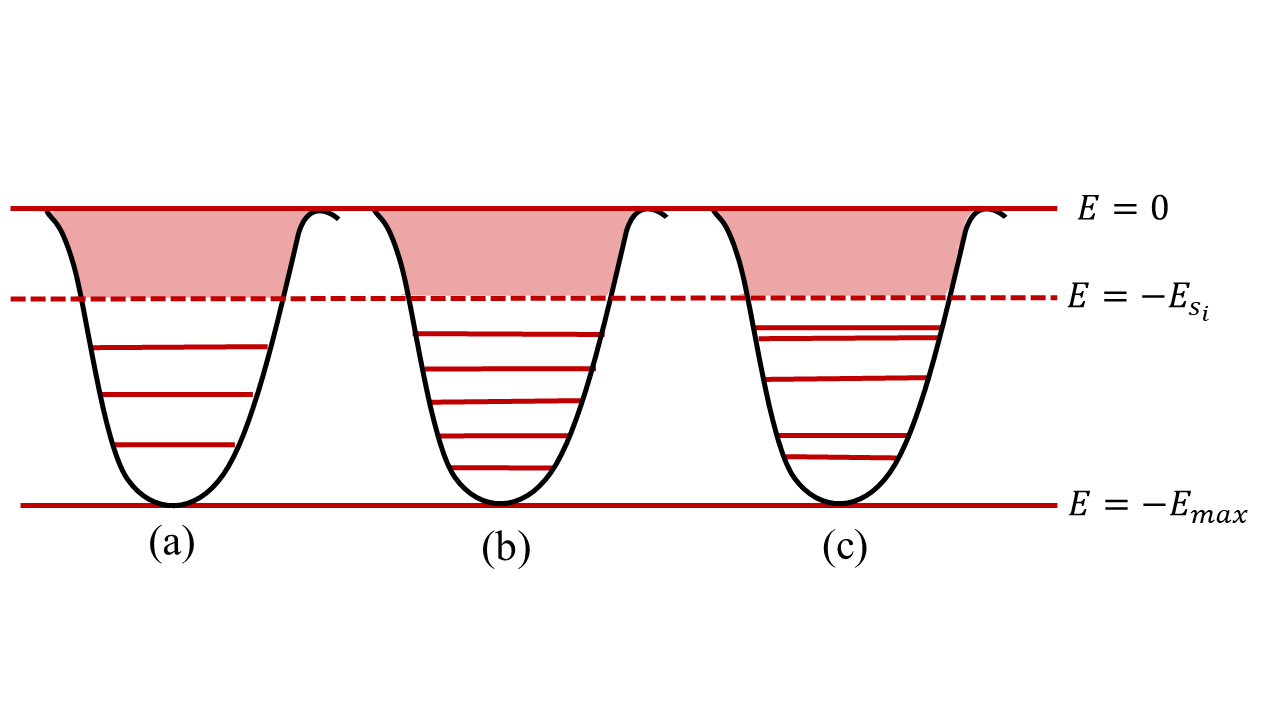}
\caption{
Illustration of energetic levels of local sites on the ordered square lattice {\bf (a)}, VRL with Model A {\bf (b)} and VRL with Model B {\bf (c)}. The minimal depth $E_{S_i}$ (no occupied neighbours) is the same for all models, as is the maximal depth $E_{max}$ (all neighbouring sites are occupied).
The disorder in the substrate  introduces additional energetic states (Model A) and also varies the gap sizes between differnt states (Model B). 
}
    \label{intutive_explanation_A_B}
\end{figure}

 The observed benefit of introduction of disorder is rooted in the definitions of transition probabilities $p_{x_i}$, $p_{x_i,x_j}$ and the local interaction energies $E_{S_i}$ and $E_{Int_i}$. 
 The interaction with the substrate, $E_{S_i}$, is the same for all lattice points in the case of ordered substrate. This is due to the fact that the cell perimeter is constant. When dealing with a VRL, each cell perimeter is different. Only on average the perimeter equals to the perimeter of the ordered lattice.
 Consequently, on average, the interaction with the substrate is the same for the ordered and the disordered cases (Model A and Model B). If one considers the situation when an adatom is isolated (step {\bf 1} in Fig.~\ref{processes}), the probability to perform a jump is similar (on average) for the different models.
 We can state that $E_{S_i}$ is the minimal "depth" of local energy trap.  On top of $E_{S_i}$ additional quantity $E_{Int_{i}}$, that depends on occupation of neighbour sites, is added.
 $E_{Int_i}$ can obtain discrete set of $4$ equally spaced values in the case of ordered square lattice. 
 This is graphically displayed in Fig.~\ref{intutive_explanation_A_B}{\bf a}, the energetic spectrum of a trap is composed of $5$ different sates. 
 When dealing with the disordered case, the situation is different for Model A and Model B. 
 First consider Model A, where $E_{Int_i}$ is defined by Eq.~(\ref{Model_A_energy}). Here the $E_{Int_i}$ also attains a set of discrete values, but the number of these values varies from site to site. For the case of $\delta=1/10$ the average number of neighbours is $6$. This means that on average, the spectrum of energetic trap in Model A is composed of $7$ different (and equally spaced) states (Fig.~\ref{intutive_explanation_A_B}{\bf b}). 
 In Model B, according to Eq.~(\ref{Model_B_energy}), there is not only $6$ distinct values (on average) but also they are not equally spaced. Fig.~\ref{intutive_explanation_A_B}{\bf c} describes the energetic spectrum of traps in Model B. In this situation one can expect to encounter with many situations where there are several energetic levels bunched closely to $E_{S_i}$.
 
 This perspective of randomization of "band gaps" and their number can help intuitively understand the benefit of introduction of disorder. At some stage the homogenization of the crystal is due to local rearrangements of the adatoms (Fig.~\ref{processes} step {\bf 4}). In order for a rearrangement to occur the adatom must escape the local energetic trap. 
 The benefit of energetic spectrum of the disordered case is due to existence of energetic states in the vicinity of $E_{S_i}$, as compared to the ordered case. This line of thought also supports the observation that the NWD value for Model B is higher as compared to Model A. Existence of "bunched" energetic states in the vicinity of $E_{S_i}$ can facilitate local transformations of adatoms and homogenize the crystal.
 We observe an effect that is small to moderate. Nonetheless, it becomes evident that disorder can  support crystal growth by facilitation of new energetic pathways.

\section{Summary} \label{my discussion}

The purpose of this work was to explore the effect of disordered substrate on crystal growth. In order to achieve this goal KMC simulations of SOS models (with disordered lattices) were performed.
The geometrical disorder of the lattice influenced the interaction energy between adatoms moving on the lattice and affected the crystal growth process. Numerically computed behavior of normalized weight density (NWD) parameter shows not only that there is a preferred temperature for crystallization but also that the presence of geometrical disorder is beneficial.
We suggest that the disorder affects the local energetic states associated with each lattice site. It creates energetic states in the close vicinity of the energy associated with free particles. Those states, that are not present in the case of ordered substrate, can produce additional (and more preferable) energetic pathways to local reformation and faster crystallization.   
It will be interesting to explore this effect further, especially since it is known that fluctuations of energetic levels can lead to effects such as stochastic resonance~\cite{Hanggi,Gitterman01}.

\bibliographystyle{apsrev4-1}
\bibliography{./deborah}

\begin{thebibliography}{42}%
\makeatletter
\providecommand \@ifxundefined [1]{%
 \@ifx{#1\undefined}
}%
\providecommand \@ifnum [1]{%
 \ifnum #1\expandafter \@firstoftwo
 \else \expandafter \@secondoftwo
 \fi
}%
\providecommand \@ifx [1]{%
 \ifx #1\expandafter \@firstoftwo
 \else \expandafter \@secondoftwo
 \fi
}%
\providecommand \natexlab [1]{#1}%
\providecommand \enquote  [1]{``#1''}%
\providecommand \bibnamefont  [1]{#1}%
\providecommand \bibfnamefont [1]{#1}%
\providecommand \citenamefont [1]{#1}%
\providecommand \href@noop [0]{\@secondoftwo}%
\providecommand \href [0]{\begingroup \@sanitize@url \@href}%
\providecommand \@href[1]{\@@startlink{#1}\@@href}%
\providecommand \@@href[1]{\endgroup#1\@@endlink}%
\providecommand \@sanitize@url [0]{\catcode `\\12\catcode `\$12\catcode
  `\&12\catcode `\#12\catcode `\^12\catcode `\_12\catcode `\%12\relax}%
\providecommand \@@startlink[1]{}%
\providecommand \@@endlink[0]{}%
\providecommand \url  [0]{\begingroup\@sanitize@url \@url }%
\providecommand \@url [1]{\endgroup\@href {#1}{\urlprefix }}%
\providecommand \urlprefix  [0]{URL }%
\providecommand \Eprint [0]{\href }%
\providecommand \doibase [0]{http://dx.doi.org/}%
\providecommand \selectlanguage [0]{\@gobble}%
\providecommand \bibinfo  [0]{\@secondoftwo}%
\providecommand \bibfield  [0]{\@secondoftwo}%
\providecommand \translation [1]{[#1]}%
\providecommand \BibitemOpen [0]{}%
\providecommand \bibitemStop [0]{}%
\providecommand \bibitemNoStop [0]{.\EOS\space}%
\providecommand \EOS [0]{\spacefactor3000\relax}%
\providecommand \BibitemShut  [1]{\csname bibitem#1\endcsname}%
\let\auto@bib@innerbib\@empty
%</preamble>
\bibitem [{\citenamefont {Novoselov}\ and\ \citenamefont
  {Neto}(2012)}]{Novoselov}%
  \BibitemOpen
  \bibfield  {author} {\bibinfo {author} {\bibfnamefont {K.~S.}\ \bibnamefont
  {Novoselov}}\ and\ \bibinfo {author} {\bibfnamefont {A.~H.~C.}\ \bibnamefont
  {Neto}},\ }\href@noop {} {\bibfield  {journal} {\bibinfo  {journal} {Physica
  Scripta}\ }\textbf {\bibinfo {volume} {T146}},\ \bibinfo {pages} {014006}
  (\bibinfo {year} {2012})}\BibitemShut {NoStop}%
\bibitem [{\citenamefont {Putkonen}\ \emph {et~al.}(2005)\citenamefont
  {Putkonen}, \citenamefont {Sajavaara}, \citenamefont {Niinist{\"o}},\ and\
  \citenamefont {Keinonen}}]{Putkonen2005}%
  \BibitemOpen
  \bibfield  {author} {\bibinfo {author} {\bibfnamefont {M.}~\bibnamefont
  {Putkonen}}, \bibinfo {author} {\bibfnamefont {T.}~\bibnamefont {Sajavaara}},
  \bibinfo {author} {\bibfnamefont {L.}~\bibnamefont {Niinist{\"o}}}, \ and\
  \bibinfo {author} {\bibfnamefont {J.}~\bibnamefont {Keinonen}},\ }\href@noop
  {} {\bibfield  {journal} {\bibinfo  {journal} {Analytical and Bioanalytical
  Chemistry}\ }\textbf {\bibinfo {volume} {382}},\ \bibinfo {pages} {1791}
  (\bibinfo {year} {2005})}\BibitemShut {NoStop}%
\bibitem [{\citenamefont {Wu}\ \emph {et~al.}(2019)\citenamefont {Wu},
  \citenamefont {Yang},\ and\ \citenamefont {Wang}}]{Wu2019}%
  \BibitemOpen
  \bibfield  {author} {\bibinfo {author} {\bibfnamefont {L.}~\bibnamefont
  {Wu}}, \bibinfo {author} {\bibfnamefont {W.}~\bibnamefont {Yang}}, \ and\
  \bibinfo {author} {\bibfnamefont {G.}~\bibnamefont {Wang}},\ }\href {\doibase
  10.1038/s41699-019-0088-4} {\bibfield  {journal} {\bibinfo  {journal} {npj 2D
  Materials and Applications}\ }\textbf {\bibinfo {volume} {3}},\ \bibinfo
  {pages} {6} (\bibinfo {year} {2019})}\BibitemShut {NoStop}%
\bibitem [{\citenamefont {Chen}\ \emph
  {et~al.}(2011{\natexlab{a}})\citenamefont {Chen}, \citenamefont {Zhu},
  \citenamefont {Chen}, \citenamefont {Z.Qiu},\ and\ \citenamefont
  {S.Jiang}}]{Chen_19}%
  \BibitemOpen
  \bibfield  {author} {\bibinfo {author} {\bibfnamefont {Z.}~\bibnamefont
  {Chen}}, \bibinfo {author} {\bibfnamefont {Y.}~\bibnamefont {Zhu}}, \bibinfo
  {author} {\bibfnamefont {S.}~\bibnamefont {Chen}}, \bibinfo {author}
  {\bibnamefont {Z.Qiu}}, \ and\ \bibinfo {author} {\bibnamefont {S.Jiang}},\
  }\href@noop {} {\bibfield  {journal} {\bibinfo  {journal} {Applied Surface
  Science}\ }\textbf {\bibinfo {volume} {257}},\ \bibinfo {pages} {6102}
  (\bibinfo {year} {2011}{\natexlab{a}})}\BibitemShut {NoStop}%
\bibitem [{\citenamefont {Nurminen}\ \emph {et~al.}(2000)\citenamefont
  {Nurminen}, \citenamefont {Kuronen},\ and\ \citenamefont
  {Kaski}}]{Nurminen_1}%
  \BibitemOpen
  \bibfield  {author} {\bibinfo {author} {\bibfnamefont {L.}~\bibnamefont
  {Nurminen}}, \bibinfo {author} {\bibfnamefont {A.}~\bibnamefont {Kuronen}}, \
  and\ \bibinfo {author} {\bibfnamefont {K.}~\bibnamefont {Kaski}},\
  }\href@noop {} {\bibfield  {journal} {\bibinfo  {journal} {Phys. Rev. B}\
  }\textbf {\bibinfo {volume} {63}},\ \bibinfo {pages} {035407} (\bibinfo
  {year} {2000})}\BibitemShut {NoStop}%
\bibitem [{\citenamefont {Tetlow}\ \emph {et~al.}(2014)\citenamefont {Tetlow}
  \emph {et~al.}}]{Tetlow_24}%
  \BibitemOpen
  \bibfield  {author} {\bibinfo {author} {\bibnamefont {Tetlow}} \emph
  {et~al.},\ }\href@noop {} {\bibfield  {journal} {\bibinfo  {journal} {Phys.
  Rep.}\ }\textbf {\bibinfo {volume} {542}},\ \bibinfo {pages} {195} (\bibinfo
  {year} {2014})}\BibitemShut {NoStop}%
\bibitem [{\citenamefont {Levi}\ and\ \citenamefont
  {Kotrla}(1997)}]{Levi_1997}%
  \BibitemOpen
  \bibfield  {author} {\bibinfo {author} {\bibfnamefont {A.~C.}\ \bibnamefont
  {Levi}}\ and\ \bibinfo {author} {\bibfnamefont {M.}~\bibnamefont {Kotrla}},\
  }\href@noop {} {\bibfield  {journal} {\bibinfo  {journal} {J. Phys.: Condens.
  Matter}\ }\textbf {\bibinfo {volume} {9}},\ \bibinfo {pages} {299} (\bibinfo
  {year} {1997})}\BibitemShut {NoStop}%
\bibitem [{\citenamefont {Jourdain}\ and\ \citenamefont
  {Bichara}(2013)}]{Jourdain}%
  \BibitemOpen
  \bibfield  {author} {\bibinfo {author} {\bibfnamefont {V.}~\bibnamefont
  {Jourdain}}\ and\ \bibinfo {author} {\bibfnamefont {C.}~\bibnamefont
  {Bichara}},\ }\href@noop {} {\bibfield  {journal} {\bibinfo  {journal}
  {Carbon}\ }\textbf {\bibinfo {volume} {58}},\ \bibinfo {pages} {2 } (\bibinfo
  {year} {2013})}\BibitemShut {NoStop}%
\bibitem [{\citenamefont {L.~Meng}\ and\ \citenamefont {Ding}(2012)}]{Meng_16}%
  \BibitemOpen
  \bibfield  {author} {\bibinfo {author} {\bibfnamefont {J.~W.}\ \bibnamefont
  {L.~Meng}, \bibfnamefont {Q~Sun}}\ and\ \bibinfo {author} {\bibfnamefont
  {F.}~\bibnamefont {Ding}},\ }\href@noop {} {\bibfield  {journal} {\bibinfo
  {journal} {J. Phys. Chem.}\ }\textbf {\bibinfo {volume} {116}},\ \bibinfo
  {pages} {6097} (\bibinfo {year} {2012})}\BibitemShut {NoStop}%
\bibitem [{\citenamefont {Shibuta}\ and\ \citenamefont
  {Maruyama}(2003)}]{Shibuta}%
  \BibitemOpen
  \bibfield  {author} {\bibinfo {author} {\bibfnamefont {Y.}~\bibnamefont
  {Shibuta}}\ and\ \bibinfo {author} {\bibfnamefont {S.}~\bibnamefont
  {Maruyama}},\ }\href@noop {} {\bibfield  {journal} {\bibinfo  {journal}
  {Chem. Phys. Lett.}\ }\textbf {\bibinfo {volume} {382}},\ \bibinfo {pages}
  {381} (\bibinfo {year} {2003})}\BibitemShut {NoStop}%
\bibitem [{\citenamefont {Zhang}\ \emph {et~al.}(2004)\citenamefont {Zhang},
  \citenamefont {Zheng}, \citenamefont {Wu}, \citenamefont {Liu},\ and\
  \citenamefont {He}}]{Zhang_17}%
  \BibitemOpen
  \bibfield  {author} {\bibinfo {author} {\bibfnamefont {P.}~\bibnamefont
  {Zhang}}, \bibinfo {author} {\bibfnamefont {X.}~\bibnamefont {Zheng}},
  \bibinfo {author} {\bibfnamefont {S.}~\bibnamefont {Wu}}, \bibinfo {author}
  {\bibfnamefont {J.}~\bibnamefont {Liu}}, \ and\ \bibinfo {author}
  {\bibfnamefont {D.}~\bibnamefont {He}},\ }\href@noop {} {\bibfield  {journal}
  {\bibinfo  {journal} {Vacuum}\ }\textbf {\bibinfo {volume} {72}},\ \bibinfo
  {pages} {405} (\bibinfo {year} {2004})}\BibitemShut {NoStop}%
\bibitem [{\citenamefont {Tang}\ \emph {et~al.}(1992)\citenamefont {Tang},
  \citenamefont {Lei-Han},\ and\ \citenamefont {Heiko}}]{Tang_25}%
  \BibitemOpen
  \bibfield  {author} {\bibinfo {author} {\bibnamefont {Tang}}, \bibinfo
  {author} {\bibnamefont {Lei-Han}}, \ and\ \bibinfo {author} {\bibfnamefont
  {L.}~\bibnamefont {Heiko}},\ }\href@noop {} {\bibfield  {journal} {\bibinfo
  {journal} {Phys. Rev. A}\ }\textbf {\bibinfo {volume} {45}},\ \bibinfo
  {pages} {R8309} (\bibinfo {year} {1992})}\BibitemShut {NoStop}%
\bibitem [{\citenamefont {Ratsch}\ and\ \citenamefont
  {Venables}(2003)}]{Ratsch_26}%
  \BibitemOpen
  \bibfield  {author} {\bibinfo {author} {\bibfnamefont {C.}~\bibnamefont
  {Ratsch}}\ and\ \bibinfo {author} {\bibfnamefont {J.}~\bibnamefont
  {Venables}},\ }\href@noop {} {\bibfield  {journal} {\bibinfo  {journal}
  {Journal of Vacuum Science and Technology A}\ }\textbf {\bibinfo {volume}
  {21}},\ \bibinfo {pages} {S96} (\bibinfo {year} {2003})}\BibitemShut
  {NoStop}%
\bibitem [{\citenamefont {Biehl}(2005{\natexlab{a}})}]{Biehl}%
  \BibitemOpen
  \bibfield  {author} {\bibinfo {author} {\bibfnamefont {M.}~\bibnamefont
  {Biehl}},\ }\href@noop {} {\bibfield  {journal} {\bibinfo  {journal}
  {International Series of Numerical Mathematics}\ }\textbf {\bibinfo {volume}
  {149}},\ \bibinfo {pages} {3–18} (\bibinfo {year}
  {2005}{\natexlab{a}})}\BibitemShut {NoStop}%
\bibitem [{\citenamefont {Battaile}\ and\ \citenamefont
  {Srolovitz}(2002)}]{dBattaile}%
  \BibitemOpen
  \bibfield  {author} {\bibinfo {author} {\bibfnamefont {C.~C.}\ \bibnamefont
  {Battaile}}\ and\ \bibinfo {author} {\bibfnamefont {D.~J.}\ \bibnamefont
  {Srolovitz}},\ }\href {\doibase 10.1146/annurev.matsci.32.012102.110247}
  {\bibfield  {journal} {\bibinfo  {journal} {Annual Review of Materials
  Research}\ }\textbf {\bibinfo {volume} {32}},\ \bibinfo {pages} {297}
  (\bibinfo {year} {2002})}\BibitemShut {NoStop}%
\bibitem [{\citenamefont {Chatterjee}\ and\ \citenamefont
  {Vlachos}(2007)}]{Chatterjee2007}%
  \BibitemOpen
  \bibfield  {author} {\bibinfo {author} {\bibfnamefont {A.}~\bibnamefont
  {Chatterjee}}\ and\ \bibinfo {author} {\bibfnamefont {D.~G.}\ \bibnamefont
  {Vlachos}},\ }\href@noop {} {\bibfield  {journal} {\bibinfo  {journal}
  {Journal of Computer-Aided Materials Design}\ }\textbf {\bibinfo {volume}
  {14}},\ \bibinfo {pages} {253} (\bibinfo {year} {2007})}\BibitemShut
  {NoStop}%
\bibitem [{\citenamefont {Pyziak}\ \emph {et~al.}(2004)\citenamefont {Pyziak},
  \citenamefont {Stefaniuk}, \citenamefont {Virt},\ and\ \citenamefont
  {Kuzma}}]{Pyziak}%
  \BibitemOpen
  \bibfield  {author} {\bibinfo {author} {\bibfnamefont {L.}~\bibnamefont
  {Pyziak}}, \bibinfo {author} {\bibfnamefont {I.}~\bibnamefont {Stefaniuk}},
  \bibinfo {author} {\bibfnamefont {I.}~\bibnamefont {Virt}}, \ and\ \bibinfo
  {author} {\bibfnamefont {M.}~\bibnamefont {Kuzma}},\ }\href {\doibase
  https://doi.org/10.1016/j.apsusc.2003.11.008} {\bibfield  {journal} {\bibinfo
   {journal} {Applied Surface Science}\ }\textbf {\bibinfo {volume} {226}},\
  \bibinfo {pages} {114 } (\bibinfo {year} {2004})}\BibitemShut {NoStop}%
\bibitem [{\citenamefont {K.}(2013)}]{Celebi_8}%
  \BibitemOpen
  \bibfield  {author} {\bibinfo {author} {\bibfnamefont {C.}~\bibnamefont
  {K.}},\ }\href@noop {} {\emph {\bibinfo {title} {Chimical vapor deposition of
  graphene on copper}}}\ (\bibinfo  {publisher} {ETH Zurich},\ \bibinfo {year}
  {2013})\BibitemShut {NoStop}%
\bibitem [{\citenamefont {Zeng}\ \emph {et~al.}(2014)\citenamefont {Zeng} \emph
  {et~al.}}]{Zeng_20}%
  \BibitemOpen
  \bibfield  {author} {\bibinfo {author} {\bibfnamefont {M.}~\bibnamefont
  {Zeng}} \emph {et~al.},\ }\href@noop {} {\bibfield  {journal} {\bibinfo
  {journal} {Chemistry of Materials}\ }\textbf {\bibinfo {volume} {26}},\
  \bibinfo {pages} {3637} (\bibinfo {year} {2014})}\BibitemShut {NoStop}%
\bibitem [{\citenamefont {Boeck}\ \emph {et~al.}(2017)\citenamefont {Boeck},
  \citenamefont {Ringleb},\ and\ \citenamefont {Bansen}}]{Boeck_27}%
  \BibitemOpen
  \bibfield  {author} {\bibinfo {author} {\bibfnamefont {T.}~\bibnamefont
  {Boeck}}, \bibinfo {author} {\bibfnamefont {F.}~\bibnamefont {Ringleb}}, \
  and\ \bibinfo {author} {\bibfnamefont {R.}~\bibnamefont {Bansen}},\
  }\href@noop {} {\bibfield  {journal} {\bibinfo  {journal} {Crystal Research
  and Technology}\ }\textbf {\bibinfo {volume} {52}},\ \bibinfo {pages}
  {1600239} (\bibinfo {year} {2017})}\BibitemShut {NoStop}%
\bibitem [{\citenamefont {Zhang}\ \emph {et~al.}(2018)\citenamefont {Zhang}
  \emph {et~al.}}]{Zhang_28}%
  \BibitemOpen
  \bibfield  {author} {\bibinfo {author} {\bibfnamefont {K.}~\bibnamefont
  {Zhang}} \emph {et~al.},\ }\href@noop {} {\bibfield  {journal} {\bibinfo
  {journal} {Proceedings of the National Academy of Sciences}\ }\textbf
  {\bibinfo {volume} {115}},\ \bibinfo {pages} {685} (\bibinfo {year}
  {2018})}\BibitemShut {NoStop}%
\bibitem [{\citenamefont {Meixner}\ \emph {et~al.}(2001)\citenamefont
  {Meixner}, \citenamefont {Sch\"oll}, \citenamefont {Shchukin},\ and\
  \citenamefont {Bimberg}}]{Meixner_23}%
  \BibitemOpen
  \bibfield  {author} {\bibinfo {author} {\bibfnamefont {M.}~\bibnamefont
  {Meixner}}, \bibinfo {author} {\bibfnamefont {E.}~\bibnamefont {Sch\"oll}},
  \bibinfo {author} {\bibfnamefont {V.~A.}\ \bibnamefont {Shchukin}}, \ and\
  \bibinfo {author} {\bibfnamefont {D.}~\bibnamefont {Bimberg}},\ }\href@noop
  {} {\bibfield  {journal} {\bibinfo  {journal} {Phys. Rev. Lett.}\ }\textbf
  {\bibinfo {volume} {87}},\ \bibinfo {pages} {236101} (\bibinfo {year}
  {2001})}\BibitemShut {NoStop}%
\bibitem [{\citenamefont {Loginova}\ \emph {et~al.}(2008)\citenamefont
  {Loginova}, \citenamefont {Bartelt}, \citenamefont {Feibelman},\ and\
  \citenamefont {McCarty}}]{Loginova_13}%
  \BibitemOpen
  \bibfield  {author} {\bibinfo {author} {\bibfnamefont {E.}~\bibnamefont
  {Loginova}}, \bibinfo {author} {\bibfnamefont {N.~C.}\ \bibnamefont
  {Bartelt}}, \bibinfo {author} {\bibfnamefont {P.~J.}\ \bibnamefont
  {Feibelman}}, \ and\ \bibinfo {author} {\bibfnamefont {K.~F.}\ \bibnamefont
  {McCarty}},\ }\href@noop {} {\bibfield  {journal} {\bibinfo  {journal} {New
  Journal of Physics}\ }\textbf {\bibinfo {volume} {10}},\ \bibinfo {pages}
  {093026} (\bibinfo {year} {2008})}\BibitemShut {NoStop}%
\bibitem [{\citenamefont {et~al.}(2013)}]{Hao_14}%
  \BibitemOpen
  \bibfield  {author} {\bibinfo {author} {\bibfnamefont {H.~Y.}\ \bibnamefont
  {et~al.}},\ }\href@noop {} {\bibfield  {journal} {\bibinfo  {journal}
  {Science}\ }\textbf {\bibinfo {volume} {342}},\ \bibinfo {pages} {720–723}
  (\bibinfo {year} {2013})}\BibitemShut {NoStop}%
\bibitem [{\citenamefont {Chen}\ \emph
  {et~al.}(2011{\natexlab{b}})\citenamefont {Chen}, \citenamefont {Tao},
  \citenamefont {Zhao},\ and\ \citenamefont {Wu}}]{Chen_18}%
  \BibitemOpen
  \bibfield  {author} {\bibinfo {author} {\bibfnamefont {H.}~\bibnamefont
  {Chen}}, \bibinfo {author} {\bibfnamefont {H.}~\bibnamefont {Tao}}, \bibinfo
  {author} {\bibfnamefont {X.}~\bibnamefont {Zhao}}, \ and\ \bibinfo {author}
  {\bibfnamefont {Q.}~\bibnamefont {Wu}},\ }\href@noop {} {\bibfield  {journal}
  {\bibinfo  {journal} {Journal of Non-Crystalline Solids}\ }\textbf {\bibinfo
  {volume} {357}},\ \bibinfo {pages} {3267 } (\bibinfo {year}
  {2011}{\natexlab{b}})}\BibitemShut {NoStop}%
\bibitem [{\citenamefont {Rafik}\ \emph {et~al.}(2012)\citenamefont {Rafik},
  \citenamefont {Arjun}, \citenamefont {Peter},\ and\ \citenamefont
  {Matthias}}]{Rafik_9}%
  \BibitemOpen
  \bibfield  {author} {\bibinfo {author} {\bibfnamefont {A.}~\bibnamefont
  {Rafik}}, \bibinfo {author} {\bibfnamefont {D.}~\bibnamefont {Arjun}},
  \bibinfo {author} {\bibfnamefont {S.}~\bibnamefont {Peter}}, \ and\ \bibinfo
  {author} {\bibfnamefont {B.}~\bibnamefont {Matthias}},\ }\href@noop {}
  {\bibfield  {journal} {\bibinfo  {journal} {App. Phys. Lett.}\ }\textbf
  {\bibinfo {volume} {100}},\ \bibinfo {pages} {021601} (\bibinfo {year}
  {2012})}\BibitemShut {NoStop}%
\bibitem [{\citenamefont {Moukarzel}\ and\ \citenamefont
  {Herrmann}(1992{\natexlab{a}})}]{Moukarzel}%
  \BibitemOpen
  \bibfield  {author} {\bibinfo {author} {\bibfnamefont {C.}~\bibnamefont
  {Moukarzel}}\ and\ \bibinfo {author} {\bibfnamefont {H.~J.}\ \bibnamefont
  {Herrmann}},\ }\href@noop {} {\bibfield  {journal} {\bibinfo  {journal} {J.
  Stat. phys.}\ }\textbf {\bibinfo {volume} {68}},\ \bibinfo {pages} {911}
  (\bibinfo {year} {1992}{\natexlab{a}})}\BibitemShut {NoStop}%
\bibitem [{\citenamefont {Bunimovich}\ and\ \citenamefont
  {Khlabystova}(2001)}]{Leonid}%
  \BibitemOpen
  \bibfield  {author} {\bibinfo {author} {\bibfnamefont {L.~A.}\ \bibnamefont
  {Bunimovich}}\ and\ \bibinfo {author} {\bibfnamefont {M.~A.}\ \bibnamefont
  {Khlabystova}},\ }\href@noop {} {\bibfield  {journal} {\bibinfo  {journal}
  {J. Stat. Phys.}\ }\textbf {\bibinfo {volume} {104}},\ \bibinfo {pages}
  {1155–1171} (\bibinfo {year} {2001})}\BibitemShut {NoStop}%
\bibitem [{\citenamefont {Schwarcz}\ \emph {et~al.}(2016)\citenamefont
  {Schwarcz}, \citenamefont {Levine}, \citenamefont {Ben-Jacob},\ and\
  \citenamefont {Ariel}}]{Schwarcz_22}%
  \BibitemOpen
  \bibfield  {author} {\bibinfo {author} {\bibfnamefont {D.}~\bibnamefont
  {Schwarcz}}, \bibinfo {author} {\bibfnamefont {H.}~\bibnamefont {Levine}},
  \bibinfo {author} {\bibfnamefont {E.}~\bibnamefont {Ben-Jacob}}, \ and\
  \bibinfo {author} {\bibfnamefont {G.}~\bibnamefont {Ariel}},\ }\href@noop {}
  {\bibfield  {journal} {\bibinfo  {journal} {Physica D: Nonlinear Phenomena}\
  }\textbf {\bibinfo {volume} {318-319}},\ \bibinfo {pages} {91} (\bibinfo
  {year} {2016})}\BibitemShut {NoStop}%
\bibitem [{\citenamefont {Moukarzel}\ and\ \citenamefont
  {Herrmann}(1992{\natexlab{b}})}]{Moukarzel_21}%
  \BibitemOpen
  \bibfield  {author} {\bibinfo {author} {\bibfnamefont {C.}~\bibnamefont
  {Moukarzel}}\ and\ \bibinfo {author} {\bibfnamefont {H.}~\bibnamefont
  {Herrmann}},\ }\href@noop {} {\bibfield  {journal} {\bibinfo  {journal} {J.
  Stat. Phys.}\ }\textbf {\bibinfo {volume} {68}},\ \bibinfo {pages} {911}
  (\bibinfo {year} {1992}{\natexlab{b}})}\BibitemShut {NoStop}%
\bibitem [{\citenamefont {Vieira}\ \emph {et~al.}(2001)\citenamefont {Vieira},
  \citenamefont {de~Carvalho},\ and\ \citenamefont
  {Salinas}}]{mixed-spin_IM_6}%
  \BibitemOpen
  \bibfield  {author} {\bibinfo {author} {\bibfnamefont {A.~P.}\ \bibnamefont
  {Vieira}}, \bibinfo {author} {\bibfnamefont {J.~X.}\ \bibnamefont
  {de~Carvalho}}, \ and\ \bibinfo {author} {\bibfnamefont {S.}~\bibnamefont
  {Salinas}},\ }\href@noop {} {\bibfield  {journal} {\bibinfo  {journal} {Phys.
  Rev. B}\ }\textbf {\bibinfo {volume} {63}},\ \bibinfo {pages} {184415}
  (\bibinfo {year} {2001})}\BibitemShut {NoStop}%
\bibitem [{\citenamefont {Biehl}(2005{\natexlab{b}})}]{michael}%
  \BibitemOpen
  \bibfield  {author} {\bibinfo {author} {\bibfnamefont {M.}~\bibnamefont
  {Biehl}},\ }\href@noop {} {\bibfield  {journal} {\bibinfo  {journal}
  {International Series of Numerical Mathematics}\ }\textbf {\bibinfo {volume}
  {149}},\ \bibinfo {pages} {3} (\bibinfo {year}
  {2005}{\natexlab{b}})}\BibitemShut {NoStop}%
\bibitem [{\citenamefont {{Xu}}\ \emph {et~al.}(2017)\citenamefont {{Xu}},
  \citenamefont {{Zapol}}, \citenamefont {{Stephenson}},\ and\ \citenamefont
  {{Thompson}}}]{Dongwei}%
  \BibitemOpen
  \bibfield  {author} {\bibinfo {author} {\bibfnamefont {D.}~\bibnamefont
  {{Xu}}}, \bibinfo {author} {\bibfnamefont {P.}~\bibnamefont {{Zapol}}},
  \bibinfo {author} {\bibfnamefont {G.~B.}\ \bibnamefont {{Stephenson}}}, \
  and\ \bibinfo {author} {\bibfnamefont {C.}~\bibnamefont {{Thompson}}},\
  }\href@noop {} {\bibfield  {journal} {\bibinfo  {journal} {\jcp}\ }\textbf
  {\bibinfo {volume} {146}},\ \bibinfo {pages} {144702} (\bibinfo {year}
  {2017})}\BibitemShut {NoStop}%
\bibitem [{\citenamefont {Bouchaud}\ and\ \citenamefont
  {Georges}(1990)}]{BOUCHAUD}%
  \BibitemOpen
  \bibfield  {author} {\bibinfo {author} {\bibfnamefont {J.-P.}\ \bibnamefont
  {Bouchaud}}\ and\ \bibinfo {author} {\bibfnamefont {A.}~\bibnamefont
  {Georges}},\ }\href@noop {} {\bibfield  {journal} {\bibinfo  {journal} {Phys.
  Rep.}\ }\textbf {\bibinfo {volume} {195}},\ \bibinfo {pages} {127 } (\bibinfo
  {year} {1990})}\BibitemShut {NoStop}%
\bibitem [{\citenamefont {Burov}\ and\ \citenamefont
  {Barkai}(2011)}]{stas2011}%
  \BibitemOpen
  \bibfield  {author} {\bibinfo {author} {\bibfnamefont {S.}~\bibnamefont
  {Burov}}\ and\ \bibinfo {author} {\bibfnamefont {E.}~\bibnamefont {Barkai}},\
  }\href@noop {} {\bibfield  {journal} {\bibinfo  {journal} {Phys. Rev. Lett.}\
  }\textbf {\bibinfo {volume} {106}},\ \bibinfo {pages} {140602} (\bibinfo
  {year} {2011})}\BibitemShut {NoStop}%
\bibitem [{\citenamefont {Miyaguchi}\ and\ \citenamefont
  {Akimoto}(2015)}]{Akimoto}%
  \BibitemOpen
  \bibfield  {author} {\bibinfo {author} {\bibfnamefont {T.}~\bibnamefont
  {Miyaguchi}}\ and\ \bibinfo {author} {\bibfnamefont {T.}~\bibnamefont
  {Akimoto}},\ }\href {\doibase 10.1103/PhysRevE.91.010102} {\bibfield
  {journal} {\bibinfo  {journal} {Phys. Rev. E}\ }\textbf {\bibinfo {volume}
  {91}},\ \bibinfo {pages} {010102} (\bibinfo {year} {2015})}\BibitemShut
  {NoStop}%
\bibitem [{\citenamefont {{Burov}}(2017)}]{stas2017}%
  \BibitemOpen
  \bibfield  {author} {\bibinfo {author} {\bibfnamefont {S.}~\bibnamefont
  {{Burov}}},\ }\href@noop {} {\bibfield  {journal} {\bibinfo  {journal}
  {\pre}\ }\textbf {\bibinfo {volume} {96}},\ \bibinfo {pages} {050103}
  (\bibinfo {year} {2017})}\BibitemShut {NoStop}%
\bibitem [{\citenamefont {Magdziarz}\ and\ \citenamefont
  {Szczotka}(0181)}]{Magdziarz2018}%
  \BibitemOpen
  \bibfield  {author} {\bibinfo {author} {\bibfnamefont {M.}~\bibnamefont
  {Magdziarz}}\ and\ \bibinfo {author} {\bibfnamefont {W.}~\bibnamefont
  {Szczotka}},\ }\href@noop {} {\bibfield  {journal} {\bibinfo  {journal} {J.
  Stat. Mech.}\ }\textbf {\bibinfo {volume} {2018}},\ \bibinfo {pages} {023207}
  (\bibinfo {year} {20181})}\BibitemShut {NoStop}%
\bibitem [{\citenamefont {Berne}\ \emph {et~al.}(1998)\citenamefont {Berne},
  \citenamefont {Ciccotti},\ and\ \citenamefont
  {Coker}}]{David_chandler_chapter}%
  \BibitemOpen
  \bibfield  {author} {\bibinfo {author} {\bibfnamefont {B.~J.}\ \bibnamefont
  {Berne}}, \bibinfo {author} {\bibfnamefont {G.}~\bibnamefont {Ciccotti}}, \
  and\ \bibinfo {author} {\bibfnamefont {D.~F.}\ \bibnamefont {Coker}},\
  }\href@noop {} {\emph {\bibinfo {title} {Classical and Quantum Dynamics in
  Condensed Phase Simulations}}}\ (\bibinfo  {publisher} {World Scientific},\
  \bibinfo {year} {1998})\ pp.\ \bibinfo {pages} {3 -- 21}\BibitemShut
  {NoStop}%
\bibitem [{\citenamefont {Nishinaga}(2015)}]{Luis}%
  \BibitemOpen
  \bibinfo {editor} {\bibfnamefont {T.}~\bibnamefont {Nishinaga}},\ ed.,\
  \href@noop {} {\emph {\bibinfo {title} {Handbook of Crystal Growth}}}\
  (\bibinfo  {publisher} {Elsevier},\ \bibinfo {year} {2015})\ pp.\ \bibinfo
  {pages} {445 -- 475}\BibitemShut {NoStop}%
\bibitem [{\citenamefont {Gammaitoni}\ \emph {et~al.}(1998)\citenamefont
  {Gammaitoni}, \citenamefont {H\"anggi}, \citenamefont {Jung},\ and\
  \citenamefont {Marchesoni}}]{Hanggi}%
  \BibitemOpen
  \bibfield  {author} {\bibinfo {author} {\bibfnamefont {L.}~\bibnamefont
  {Gammaitoni}}, \bibinfo {author} {\bibfnamefont {P.}~\bibnamefont
  {H\"anggi}}, \bibinfo {author} {\bibfnamefont {P.}~\bibnamefont {Jung}}, \
  and\ \bibinfo {author} {\bibfnamefont {F.}~\bibnamefont {Marchesoni}},\
  }\href {\doibase 10.1103/RevModPhys.70.223} {\bibfield  {journal} {\bibinfo
  {journal} {Rev. Mod. Phys.}\ }\textbf {\bibinfo {volume} {70}},\ \bibinfo
  {pages} {223} (\bibinfo {year} {1998})}\BibitemShut {NoStop}%
\bibitem [{\citenamefont {Burov}\ and\ \citenamefont
  {Gitterman}(2016)}]{Gitterman01}%
  \BibitemOpen
  \bibfield  {author} {\bibinfo {author} {\bibfnamefont {S.}~\bibnamefont
  {Burov}}\ and\ \bibinfo {author} {\bibfnamefont {M.}~\bibnamefont
  {Gitterman}},\ }\href {\doibase 10.1103/PhysRevE.94.052144} {\bibfield
  {journal} {\bibinfo  {journal} {Phys. Rev. E}\ }\textbf {\bibinfo {volume}
  {94}},\ \bibinfo {pages} {052144} (\bibinfo {year} {2016})}\BibitemShut
  {NoStop}%
\end{thebibliography}%
\end{document}